\documentclass[12pt,a4paper]{article}

\usepackage{graphicx}  
\usepackage{xspace}
\usepackage{color}
\usepackage{colortbl}

\usepackage{amsmath}

\usepackage{ifthen} 
\newboolean{pdflatex}
\setboolean{pdflatex}{true} 

\usepackage{rotating}

\newboolean{uprightparticles}
\setboolean{uprightparticles}{false} 
\usepackage{amssymb}
\usepackage{amsfonts}
\usepackage{upgreek}
\usepackage{lhcb-symbols-def}

\begin{document}


\begin{titlepage}
\pagenumbering{roman}

\vspace*{-1.5cm}
\centerline{\large EUROPEAN ORGANIZATION FOR NUCLEAR RESEARCH (CERN)}
\vspace*{1.5cm}
\hspace*{-0.5cm}
\begin{tabular*}{\linewidth}{lc@{\extracolsep{\fill}}r}
\ifthenelse{\boolean{pdflatex}}
\vspace*{-2.7cm}\mbox{\!\!\!\includegraphics[width=.14\textwidth]{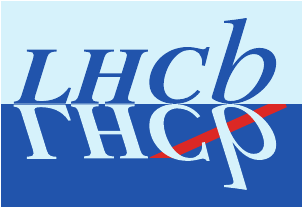}}
& & \\
 & & LHCb-PAPER-2011-010 \\  
 &  & CERN-PH-EP-2011-194 \\
 & & \today \\ 
 & & \\
\end{tabular*}

\vspace*{4.0cm}

{\bf\boldmath\huge
\begin{center}
Measurement of the \Bs-\Bsb oscillation frequency \dms in \BsDpithreepi decays
\end{center}
}
\begin{center}
Accepted by Phys. Lett. B
\end{center}
\vspace*{2.0cm}

\vspace{\fill}

\begin{abstract}
  \noindent
The \Bs-\Bsb oscillation frequency \dms is
measured with 36 pb$^{-1}$ of data collected in $pp$ collisions at $\sqrt{s}$
= 7 TeV by the \lhcb experiment at the Large Hadron Collider.
A total of  1381 \BsDpi and \BsDthreepi signal decays are reconstructed, with  average decay time resolutions of 
44~fs and 36~fs, respectively. An oscillation signal with a statistical significance of
4.6\,$\sigma$ is observed. The measured oscillation frequency is \dms = 17.63
$\pm$ 0.11 (stat) $\pm$ 0.02 (syst)~ps$^{-1}$.  
\end{abstract}

\end{titlepage}


\newpage
\setcounter{page}{2}
\mbox{~}\\
\newpage

\mbox{~}\\
\vspace{-1.5cm}

\mbox{~}\\
R.~Aaij$^{23}$, 
B.~Adeva$^{36}$, 
M.~Adinolfi$^{42}$, 
C.~Adrover$^{6}$, 
A.~Affolder$^{48}$, 
Z.~Ajaltouni$^{5}$, 
J.~Albrecht$^{37}$, 
F.~Alessio$^{37}$, 
M.~Alexander$^{47}$, 
G.~Alkhazov$^{29}$
% ...
\\
{\it \footnotesize
$ ^{1}$Centro Brasileiro de Pesquisas F\'{i}sicas (CBPF), Rio de Janeiro, Brazil\\
% ...
$ ^{5}$Clermont Universit\'{e}, Universit\'{e} Blaise Pascal, CNRS/IN2P3, LPC, Clermont-Ferrand, France\\
$ ^{6}$CPPM, Aix-Marseille Universit\'{e}, CNRS/IN2P3, Marseille, France\\
% ...
$ ^{23}$Nikhef National Institute for Subatomic Physics, Amsterdam, Netherlands\\
% ...
$ ^{29}$Petersburg Nuclear Physics Institute (PNPI), Gatchina, Russia\\
% ...
$ ^{36}$Universidad de Santiago de Compostela, Santiago de Compostela, Spain\\
$ ^{37}$European Organization for Nuclear Research (CERN), Geneva, Switzerland\\
% ...
$ ^{42}$H.H. Wills Physics Laboratory, University of Bristol, Bristol, United Kingdom\\
% ...
$ ^{47}$School of Physics and Astronomy, University of Glasgow, Glasgow, United Kingdom\\
$ ^{48}$Oliver Lodge Laboratory, University of Liverpool, Liverpool, United Kingdom\\
% ...
}


\clearpage

\pagestyle{plain} 
\setcounter{page}{1}
\pagenumbering{arabic}

\clearpage
\section{Introduction}
After the observation of \Bd-\Bdb mixing and the measurement of its strength in 1987 \cite{bib:argus}, 
it took a further 19 years for the \Bs-\Bsb
frequency to be measured for the first time \cite{bib:bsmixingd0},\cite{bib:bsmixing}. 
This is mainly due to the fact that the 
\Bs-\Bsb
oscillation frequency is 35 times larger than that for the \Bd-\Bdb system, 
posing a considerable challenge for the decay time
resolution of detectors. For the LHCb experiment, the ability to
resolve these fast \Bs-\Bsb oscillations
is a prerequisite for many physics analyses. 
In particular it is essential for the study of the 
time-dependent \CP asymmetry of \BsToJPsiPhi decays
 \cite{bib:niersteLenz}. 
The oscillation frequency in the \Bs-\Bsb system is given by the mass
difference between the heavy and light mass eigenstates, $\Delta m_s$ (we use
units with $\hbar$\,=\,1). 
In this letter, we
report a measurement of \dms by the LHCb experiment with data
collected in 2010.

The LHCb spectrometer covers the pseudo-rapidity range
2 to 5. 
In this region, $b$ hadrons are produced with a large Lorentz boost and have an 
average flight path of 7~mm. The LHCb detector consists of several components
arranged along the LHC beam line. The
vertex detector (VELO) surrounds the collision point, followed by a
first Ring Imaging Cherenkov (RICH) counter, a tracking station,
a dipole magnet, three more tracking stations, a second RICH detector, a
calorimeter system and a muon detector. The calorimeter system consists of a
scintillating pad detector (SPD), 
a preshower detector, an electromagnetic calorimeter and a hadronic
calorimeter.
A detailed description of the
detector can be found in Ref.~\cite{bib:lhcb_detector}. 
The precise spatial resolution of the VELO results in an impact parameter resolution of 20--50 $\mu$m
 in the $x$ and $y$ directions\footnote{LHCb uses a right-handed Cartesian
   coordinate system
with the $x$ direction pointing inside the LHC ring, the $y$ direction
pointing upwards and the $z$ direction running along the beamline from the
interaction point towards the spectrometer.} for charged particles with transverse momenta in the
 range relevant
 for \Bs daughter tracks used in this analysis. The $x$ and $y$ resolution in
 the position of
 the primary vertex reconstruction is about
 15~$\mu$m while the $z$ resolution is about 80 $\mu$m. 
This excellent performance results in the decay time
 resolution needed to observe the fast \Bs-\Bsb oscillations.
The invariant mass
resolution provided by the tracking system and the \pion/\kaon separation
given by the two RICH detectors provide clean \Bs meson signals with
small background. 
The particle identification capabilities of the RICH together with the
calorimeter and muon systems allow the initial flavour of the \Bs to be
tagged using  charged kaons, electrons and muons, respectively.

In the next section, the data sample used 
and the analysis strategy are introduced. This is followed by descriptions
of the analysis of the invariant mass and decay time
distributions, and the flavour tagging. Finally, we discuss
the fit result for the oscillation frequency and the associated
systematic uncertainties.

\section{Data sample and analysis strategy}
The analysis uses \Bs 
candidates reconstructed in four flavour-specific decay modes,\footnote{Unless
  explicitly stated, inclusion of charge-conjugated modes
  is implied.}
namely \BsDphipipiwoall, \BsDKStKpiwoall, \BsDKKpipiwo and
\BsDthreepiwoall.
To avoid double counting, candidates that pass the selection
criteria of one mode are not considered for the following modes.
All reconstructed decays are flavour-specific final states, thus the flavour of
the \Bs at the time of its decay is given by the charges of the final state
 particles of the decay. 
A combination of tagging algorithms is used to identify the \Bs
flavour at production. The algorithms provide for each event a tagging decision as well as an
estimate of the probability that this decision is wrong (mistag
probability). These algorithms have been optimized and calibrated using
large event samples of flavour-specific $B \rightarrow \mu^+ D^{*-} X$ and
\BuJK decays and a sample of \BdDPi decays.

The analysis is based on a data set of
36~pb$^{-1}$ of $pp$ collisions at $\sqrt{s} = 7$~TeV collected in
2010.  The first trigger level is
implemented in hardware, while the second trigger level is
based on software. 
Trigger conditions were progressively tightened over the
duration of the data taking period to cope with the rapidly increasing
instantaneous
luminosities delivered by the LHC.
In the hardware trigger, the events
 used in this analysis were selected by requiring 
a cluster with a minimum transverse energy in the hadronic calorimeter. The
applied threshold was increased from 2.5 to 3.6 \gev throughout the data
taking period. A cut on
the number of hits in the SPD detector was applied to reject very high occupancy events. 
The software
trigger for the first 2.4~pb$^{-1}$ of data required a good quality
displaced vertex reconstructed from two tracks with transverse momenta $p_{\rm T}$ of
at least 500~\mevc. For the remaining data, a
two-level software trigger was applied. A good quality track
with large impact parameter with respect to the primary vertex was required 
with $p_{\rm T}>1.85$~\gevc and momentum $p>13.3$~\gevc \cite{HLT1}. For 
  events passing these criteria, a good quality displaced vertex was required,
  formed out of two tracks with $p_{\rm T}>0.5$~\gevc and
  $p>5$~\gevc and with a mass variable in the range 2 to
  7~\gevcc \cite{HLT}.

Some of the offline event selection criteria are optimized individually for 
each of the four decay modes
under study. In this way specific features such as the
masses of the intermediate $\phi$ and $K^{*0}$ resonances or 
the helicity angle distribution of the $K^{*0}$ can be used.  
The selection criteria common to all decay
modes exploit the long \Bs lifetime by
applying cuts on the impact parameters of the daughter tracks, on the angle
of the reconstructed \Bs momentum relative to the line between the
reconstructed primary vertex and the \Bs vertex 
and on the \Bs decay time. 
Additional cuts are applied on the $p$ and $p_T$ of 
the \Bs candidate and its decay products
  as well as on particle identification variables and on track and vertex
  quality. Finally, cuts on the impact parameter significance of the
  reconstructed \Dsm and its distance of closest approach to the primary vertex
  are applied.
The reconstructed \Dsm mass is required to be consistent with the PDG value~\cite{bib:PDG}.
After this selection, a total of about 14,400 candidates remain in the \BsDpi  invariant mass
window of [4.80,~5.85]~\gevcc and in the \BsDthreepi invariant mass window of [5.00,~5.60]~\gevcc.

An unbinned likelihood method is employed to fit simultaneously 
the invariant mass and decay time distributions of the
four decay modes. The probability density
functions (PDFs) for the signal and for the background in each of the
four modes can be written as

\begin{equation}
\mathcal{P} = \mathcal{P}_m(m) \, \mathcal{P}_t(t, q | \sigma_t, \eta) 
\, \mathcal{P}_{\sigma_t}(\sigma_t) \, \mathcal{P}_{\eta}(\eta),
\end{equation}
where $m$ is the reconstructed invariant mass of the \Bs candidate, $t$
is its reconstructed decay time and $\sigma_t$ is the event-by-event
estimate of the decay time resolution given by the event
reconstruction algorithm. The tagging decision $q$ can be 0 (no tag), $-1$
(different flavour at production and decay) or $+1$ (same flavour at
production and decay). The predicted event-by-event mistag probability $\eta$
can take values between 0 and 0.5. 
The terms $\mathcal{P}_m$ and $\mathcal{P}_t$ describe the invariant
mass distribution and the decay time distribution, respectively. $\mathcal{P}_t$ is a conditional probability depending on
$\sigma_t$ and $\eta$. The terms $\mathcal{P}_{\sigma_t}$ and $\mathcal{P}_{\eta}$ are
required to ensure the proper relative
normalization of $\mathcal{P}_t$ for signal and background~\cite{bib:Punzi}. These terms
are determined directly from the data, using the measured distribution
in the upper $B^0_s$ invariant mass sideband for the background PDF and the
sideband subtracted distribution in the invariant mass signal region
for the signal PDF.

\section{Fit to the invariant mass distributions}
\begin{figure}[tb]
  \begin{center}
        \includegraphics[angle=0, width=0.41\textwidth, height=5cm]{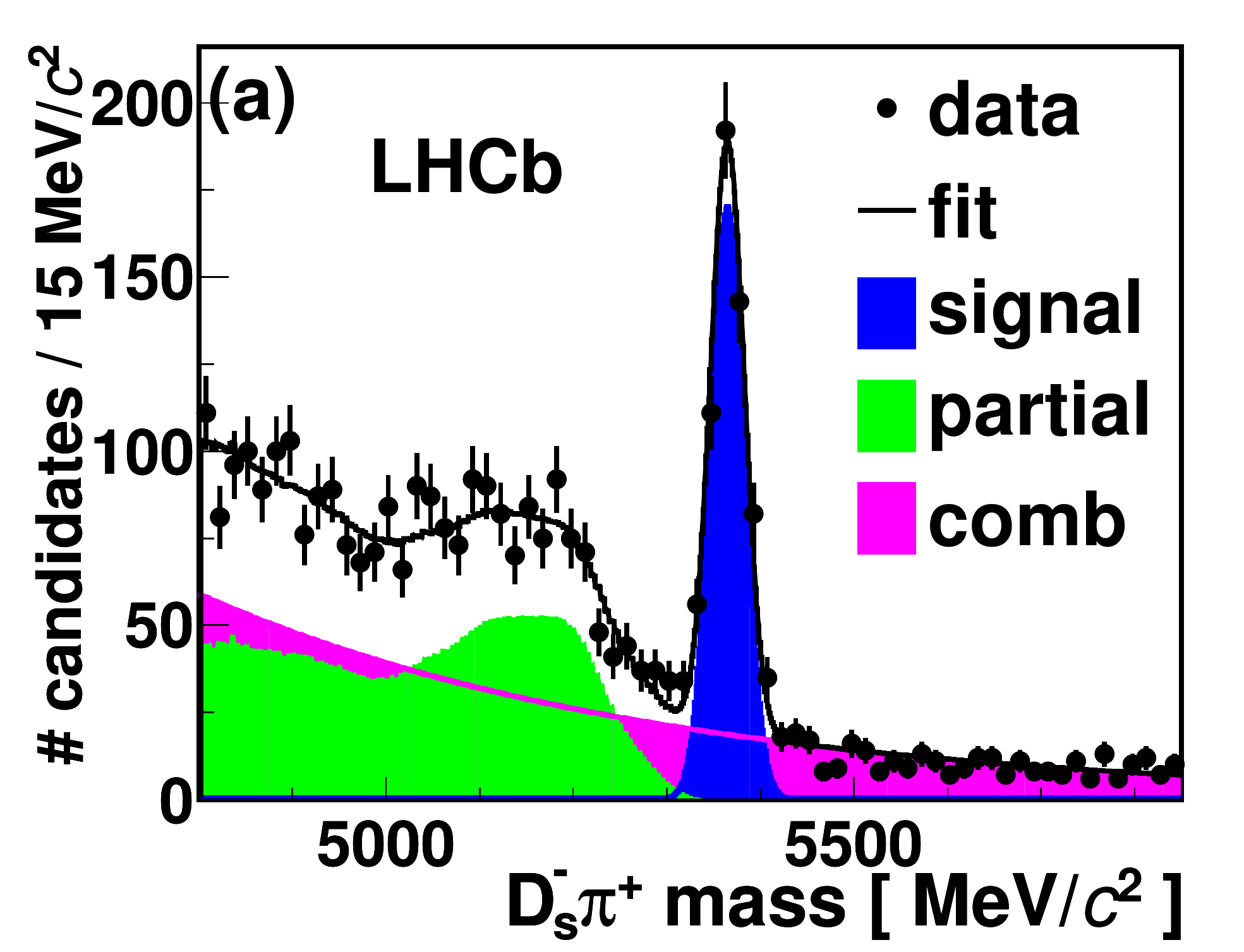}
        \includegraphics[angle=0, width=0.41\textwidth, height=5cm]{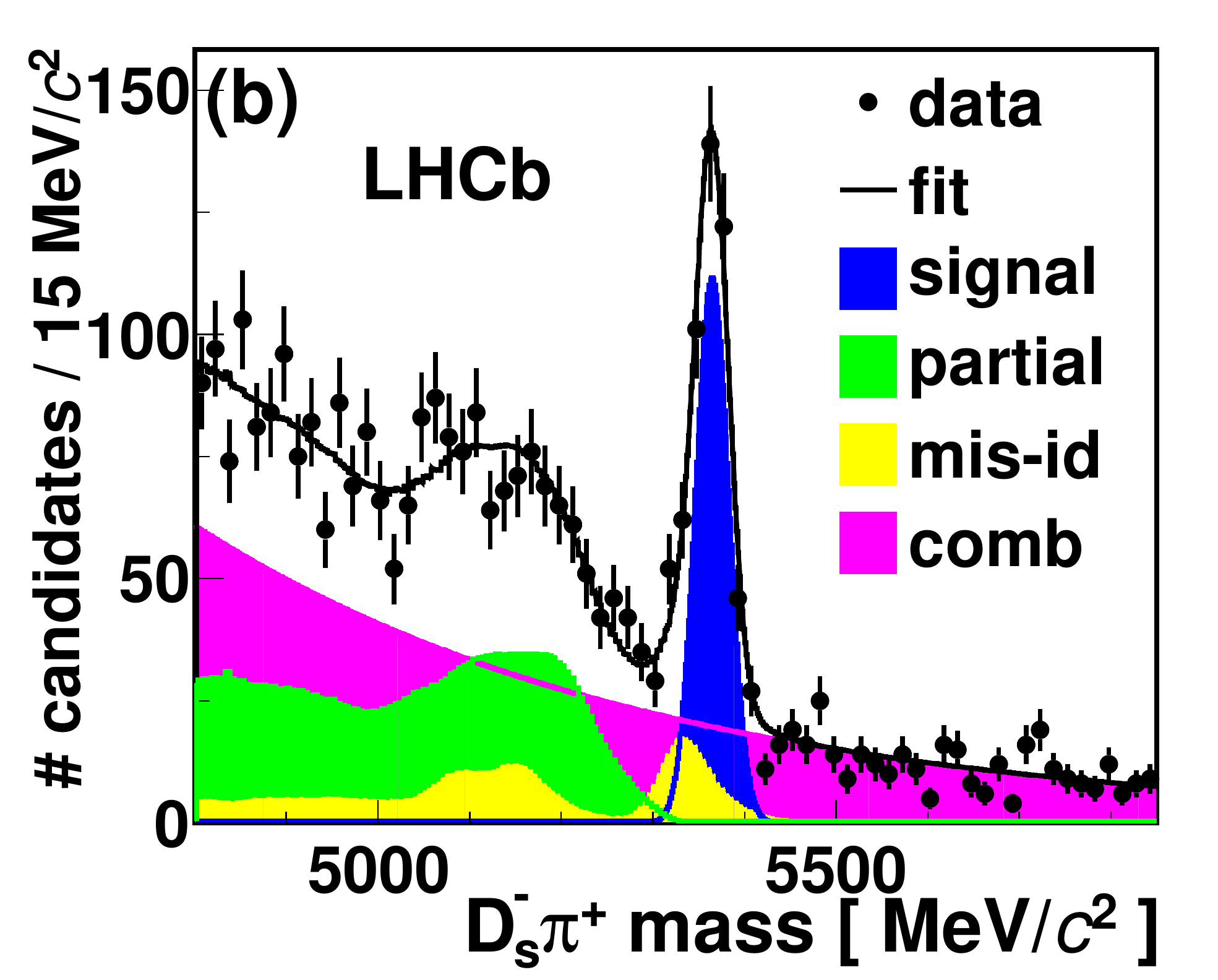}\\
        \includegraphics[angle=0, width=0.41\textwidth, height=5cm]{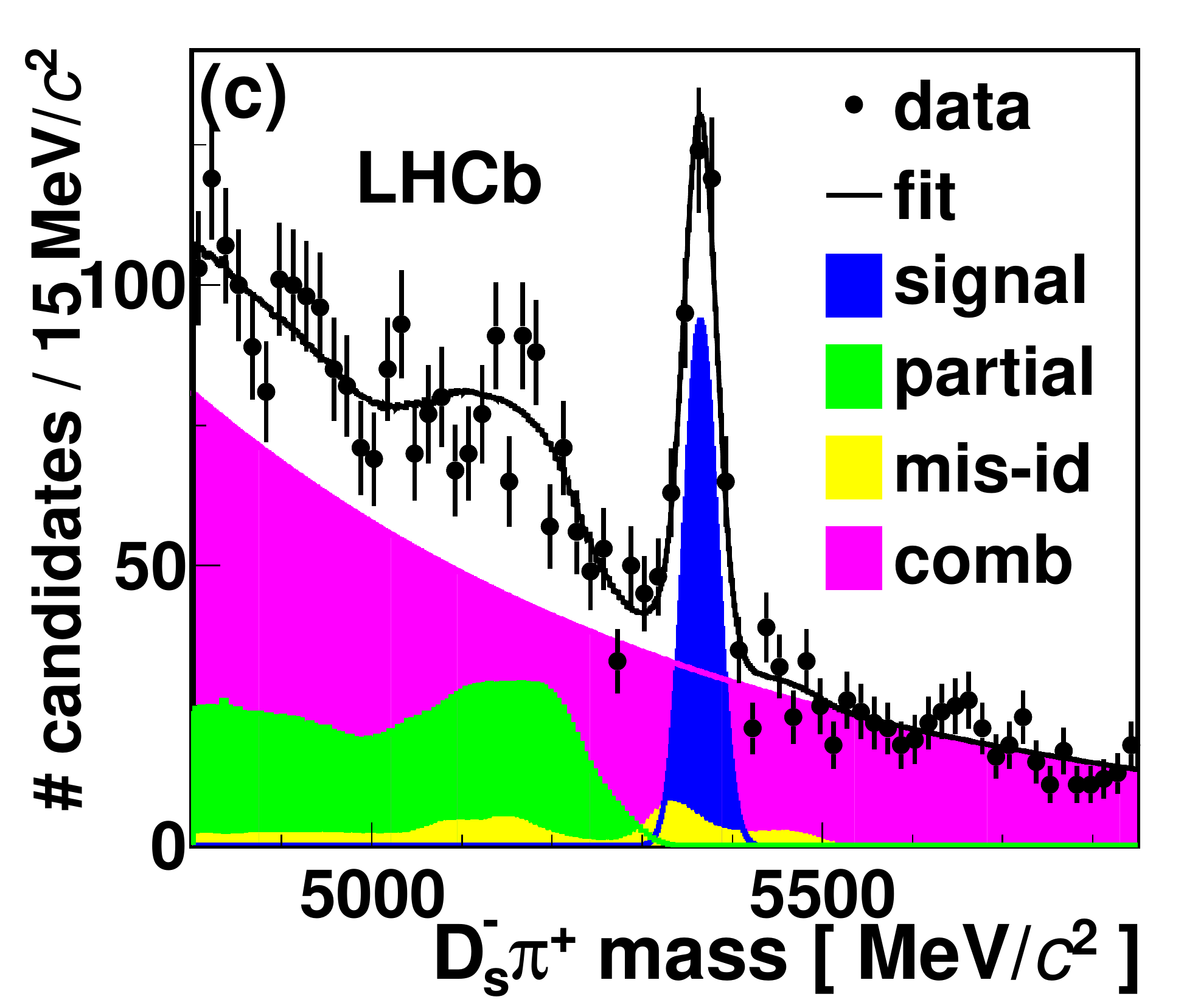}
        \includegraphics[angle=0, width=0.41\textwidth, height=5cm]{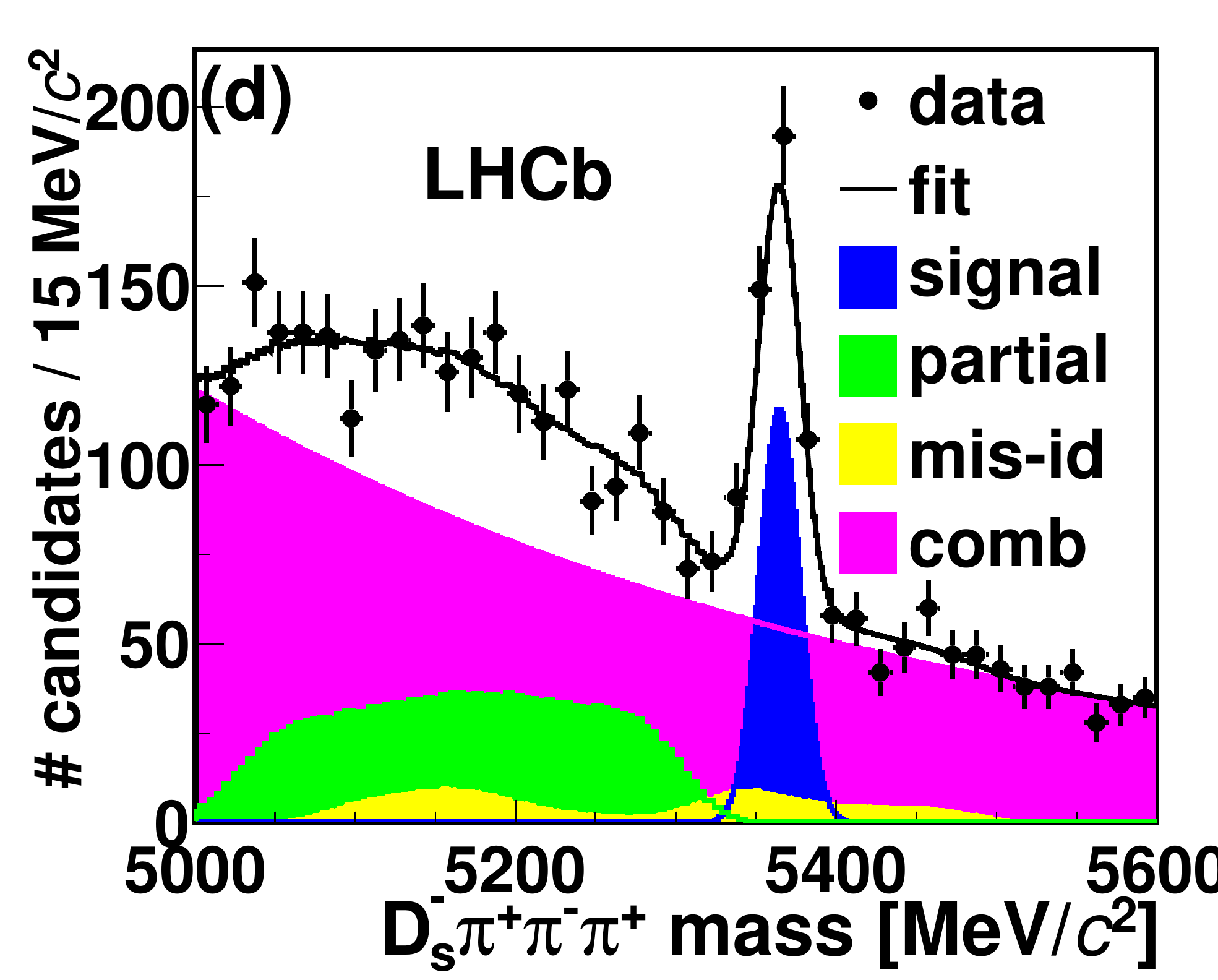}\\
        \caption{Mass distributions for
          a)~\BsDphipipi, b)~\BsDKStKpi, c)~\BsDKKpipi and d)~\BsDthreepi candidates.
          The fits
          and the various background components are described in the text.
          ``Partial''
          refers to background from partially reconstructed $B^0_s$ decays,
          ``mis-id''  refers to background from fully or partially reconstructed
          $B^0$ and $\Lambda_b$ decays
          with one mis-identified daughter particle, and ``comb'' refers to
          combinatorial background.}
    \label{fig:Bsmass}
  \end{center}
\end{figure}
The invariant mass of each \Bs candidate is determined in a vertex fit using
a constraint on the \Dsm mass. The invariant mass 
spectra for the four decay modes after all selection criteria
are shown in Fig.\ \ref{fig:Bsmass}. 
The four
  distributions are fit simultaneously taking into
  account contributions from signal, combinatorial background and
  $b$ decay backgrounds. 
  The signals are described by Gaussian distributions. The fit constrains
  the mean of the Gaussian distributions to be
  the same for all four decay modes, whereas it allows the width to be 
  different for the \BsDpi and the \BsDthreepi modes, respectively.
  The combinatorial
  backgrounds are described by exponential functions. Their parameters 
  are allowed to vary individually for the four decay modes. An alternative 
  parameterization of the combinatorial backgrounds by a first order
  polynomial is used as part of the systematic studies.

The $b$ decay backgrounds include partially reconstructed
  $B^0_s$ decays, as well as fully and partially reconstructed \Bd and
$\Lambda_b$ decays with one mis-identified daughter
  particle.
Their shapes
  are derived from a large simulated event sample, where all
  selection cuts were applied on generator level quantities. The 
  invariant mass spectra were then smeared with a Gaussian distribution to take
  into account effects of detector resolution. This approach was
  validated by comparing the results with those from a full 
   simulation including a detailed description of the detector response. 
  The relative normalization factors for the different $b$ decay
  backgrounds are parameters in the fit. 
They are constrained to be the same for the three \BsDpi decay modes.

  The fit returns a value of $m(B^0_s)$ = 5364.7 $\pm$ 0.7 \mevcc,
  about 1.5 \mevcc below the PDG value \cite{bib:PDG}. This mass shift is 
  attributed to imperfections in the  detector
  alignment and magnetic field calibration. A dedicated study on the momentum
  scale resulted in
  a correction for this effect \cite{Bmass}. This calibration procedure is however not used for the
  analysis presented here as the momentum scale correction largely cancels in the
  calculation of \dms. The mass templates describing $b$ decay backgrounds
  are shifted according to the observed bias. The fit gives signal mass
  resolutions of $\sigma_m$ = 18.1 \mevcc for the \BsDpi
  modes and $\sigma_m$ = 12.7 \mevcc for the \BsDthreepi
  mode, respectively. The signal yields extracted from the fit are summarized in
  Table\ \ref{tab:signal}. For the remainder of the analysis, the 
  invariant $B^0_s$ mass range is limited to [$m(\Bs) -3\sigma_m$, 5.85~GeV/$c^2$] and [$m(\Bs)
  -3\sigma_m$, 5.60 GeV/$c^2$] for the \BsDpi and \BsDthreepi modes,
  respectively. 
The lower cut of this asymmetric  mass window is chosen to reject all background candidates from partial
reconstructed $B^0_s$ decays. The only remaining $b$ decay backgrounds are thus due to
mis-identified \Bd and $\Lambda_b$ decays. The candidates in the high mass sidebands provide a
clean sample of combinatorial background. Including them in the fit permits 
to determine the decay time distribution and tagging behaviour of
this background contribution. \\
  The parameters derived in the fit to the mass distributions are
fixed for the remainder of the analysis.

\begin{table}[tb]
\begin{center}
\caption{$B^0_s$ signal yields. \label{tab:signal}}
\vspace{1mm}

\begin{tabular}{|l|c|} \hline
Decay mode     & Signal yield \\ \hline
\BsDphipipi    & 515 $\pm$ 25 \\
\BsDKStKpi     & 338 $\pm$ 27 \\
\BsDKKpipi     & 283 $\pm$ 27 \\
\BsDthreepi    & 245 $\pm$ 46 \\ \hline
Total          & 1381 $\pm$ 65~ \\ \hline
\end{tabular}
\end{center}
\end{table}

\section{Fit to the decay time distribution}
Ignoring detector resolution effects, selection biases and flavour tagging, 
the distribution of the decay time $t$ of the signal is described by

\begin{equation}
\mathcal{P}_t(t) \propto \Gs\, e^{-\Gs \, t } \, \cosh \left ( \frac{\Delta
\Gamma_s}{2}t \right ) \, \theta(t), \label{eq:proptime} 
\end{equation}
where \Gs\xspace is the \Bs decay width and \DGs\xspace the decay width
difference between the heavy and the light mass eigenstates. In the fit
\DGs\xspace is fixed to its PDG value of 0.09\,\Gs~\cite{bib:PDG}. 
As part of
the evaluation of systematic uncertainties on $\Delta m_s$, the assumed
value of \DGs~is varied within its current uncertainty
between 0 and 0.2\,\Gs.
The step function $\theta(t)$ restricts the PDF to positive decay times. 

The true decay time is convolved with the decay time resolution function of
the detector.
An event-by-event estimate of the decay time  resolution is calculated by the fitting algorithm, which reconstructs the
decay vertex of the \Bs and computes its
decay length and decay time. 
No constraint on the
\Dsm mass is applied in the computation of the decay time in order to
minimize sensitivity to the knowledge of the momentum scale of the
experiment.
The decay time uncertainty calculated by the fitting algorithm does not include
possible effects from an imperfect understanding of the detector
material or its spatial alignment. To correct for such effects, the
calculated event-by-event decay time uncertainties, $\sigma_{t}$, are multiplied by a
constant scale factor $S_{\sigma_t}$. The value of $S_{\sigma_t}$ is determined from
data, using a sample of fake \Bs candidates formed by a prompt \Dsm and a
\pip from the primary vertex. The contamination due to secondary \Dsm from
$B$ decays is estimated and statistically subtracted using the
measured \Dsm impact parameter distribution. The distribution of decay
times for this fake $B^0_s$ sample, each divided by its calculated
event-by-event uncertainty, is fitted with a Gaussian function and
$S_{\sigma_t}$ is taken as the resulting standard deviation.
Using the full sample of fake \Bs candidates, a value of
$S_{\sigma_t}$=1.3 is obtained. This value is used as the nominal scale factor in the \dms
analysis. Studying different regions of phase space of the fake \Bs candidates
separately, values for $S_{\sigma_t}$ between 1.2 and 1.4 are obtained. This
variation is taken into account for evaluating the systematic uncertainties on \dms.
Including the nominal scale factor $S_{\sigma_t}$=1.3, the average decay time
resolution is 44~fs for the \BsDpi sample and 36~fs for
the \BsDthreepi sample. The decay time resolution
is taken into account in the PDF by convolving Eq.
\ref{eq:proptime} with a Gaussian $G$ with mean zero and standard deviation
1.3\,$\sigma_t$.

\begin{figure}[tb]
  \begin{center}
        \includegraphics[angle=0, width=0.41\textwidth, height=5.5cm]{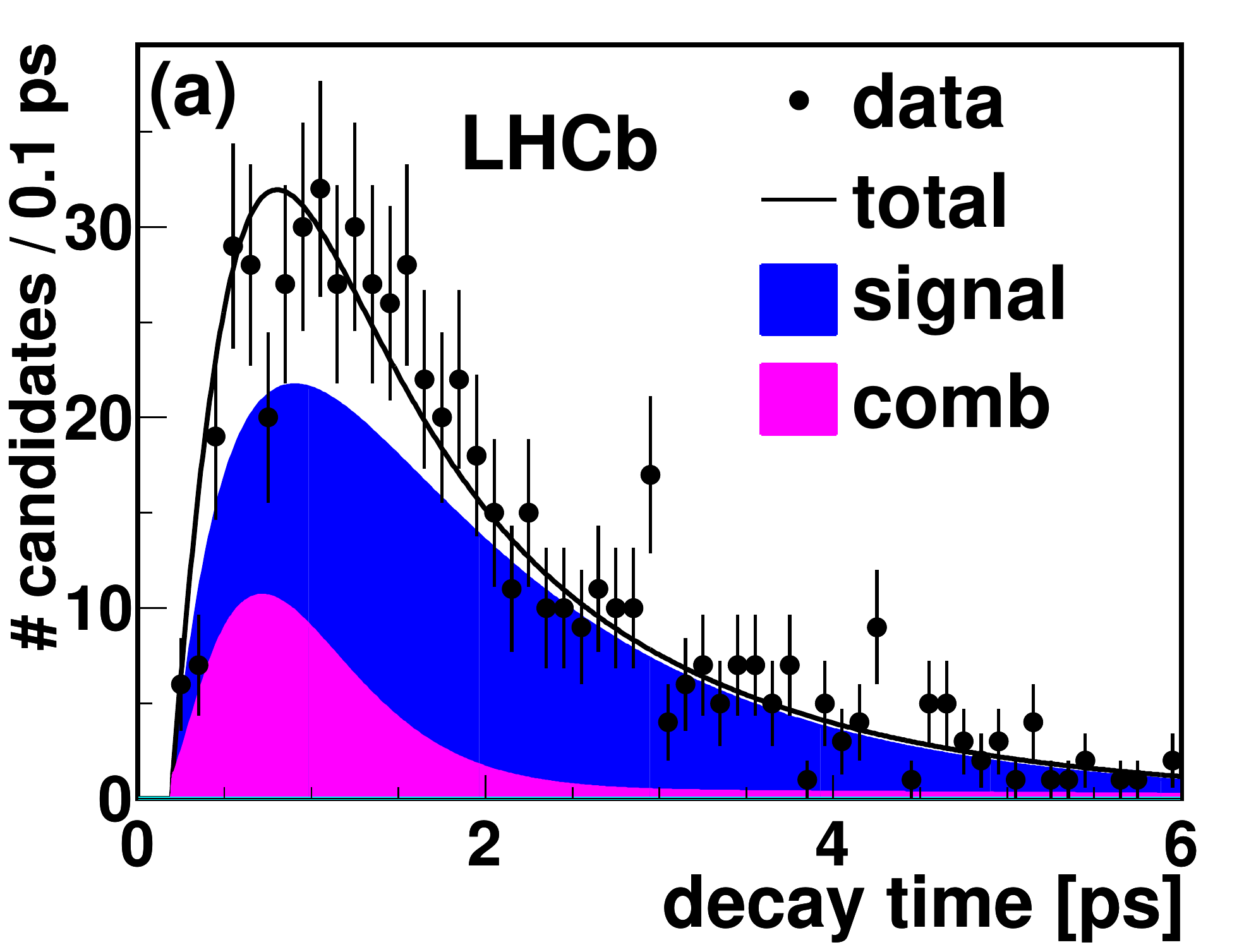}
        \includegraphics[angle=0,width=0.41\textwidth, height=5.5cm]{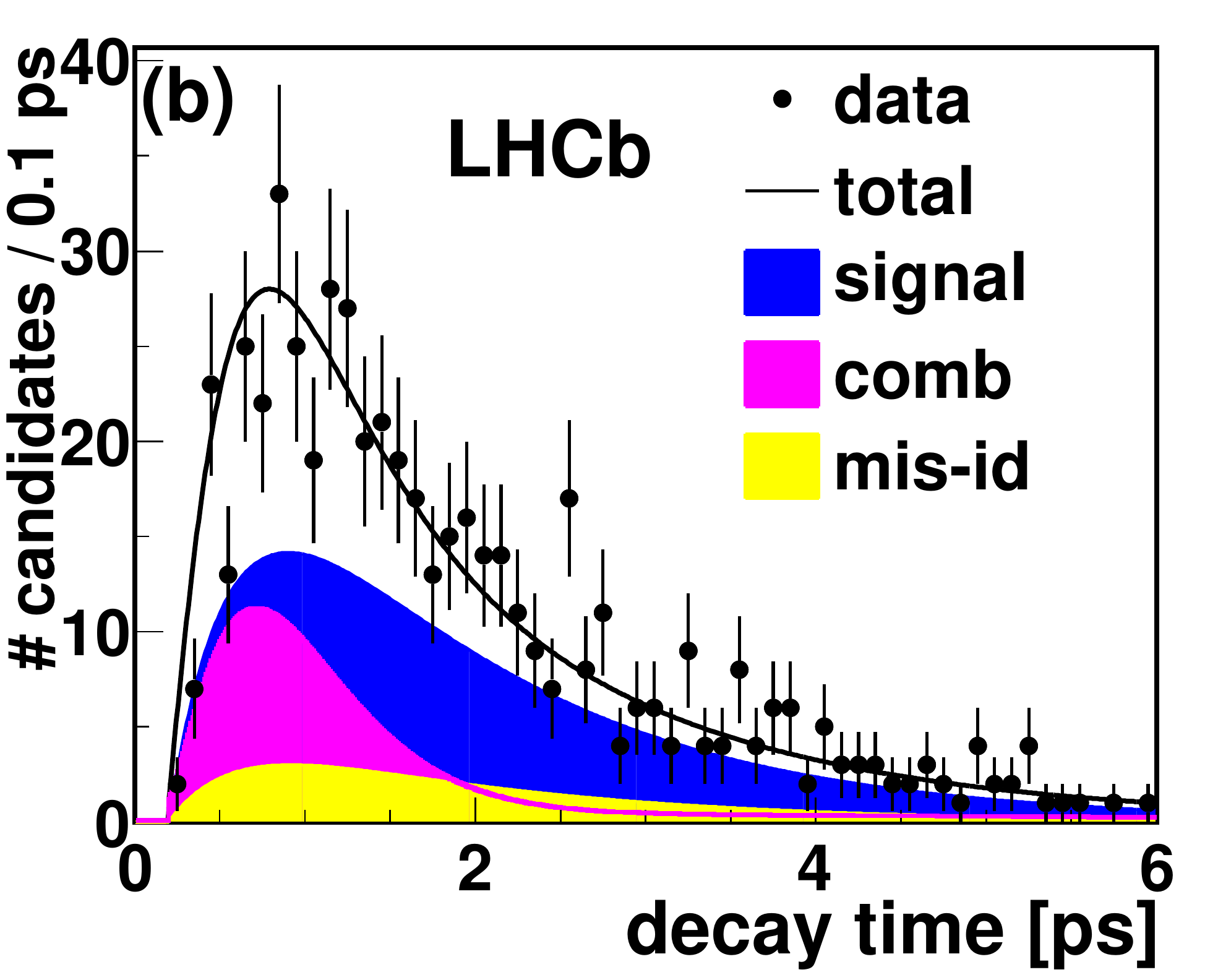}\\
        \includegraphics[angle=0, width=0.41\textwidth, height=5.5cm]{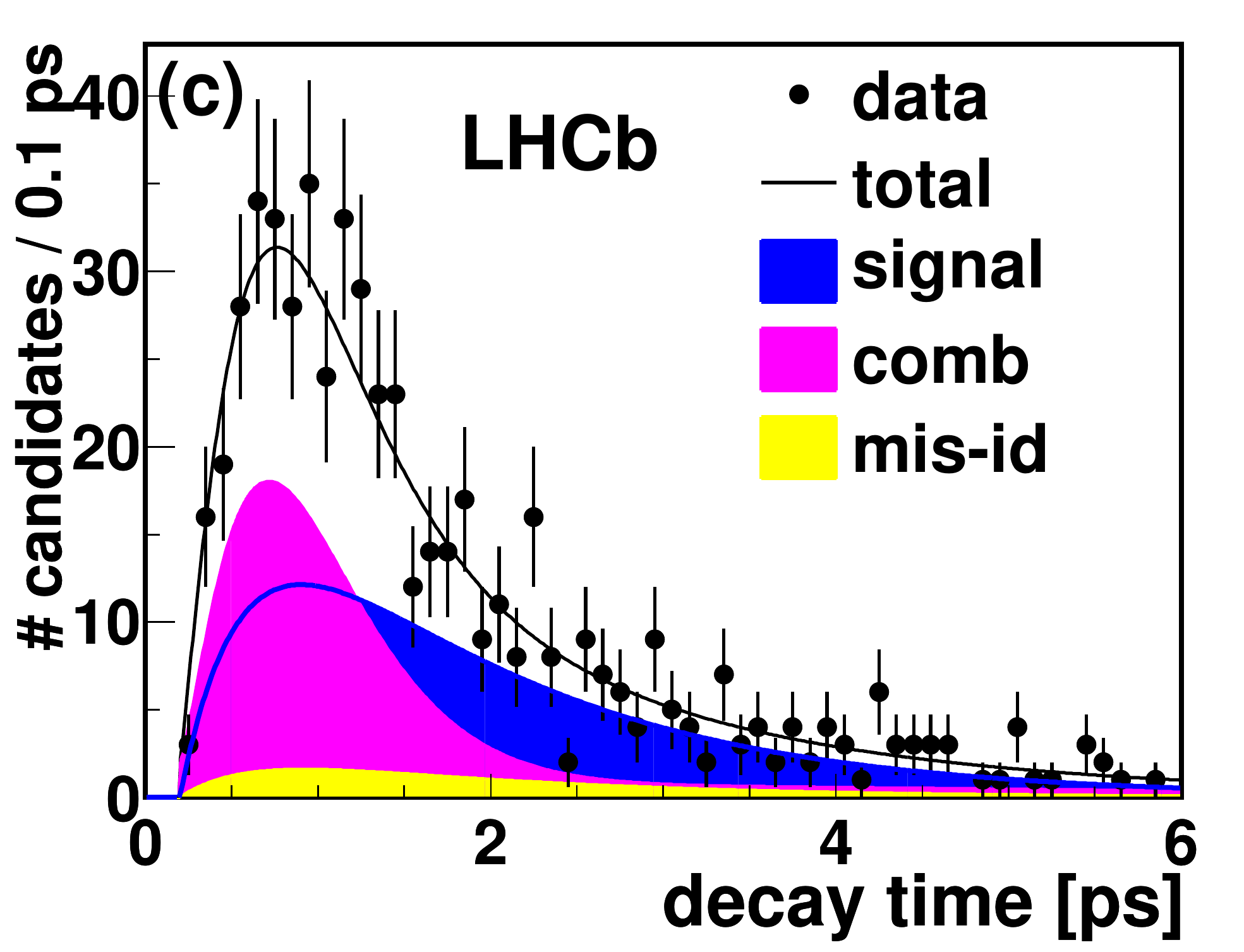}
        \includegraphics[angle=0, width=0.41\textwidth, height=5.5cm]{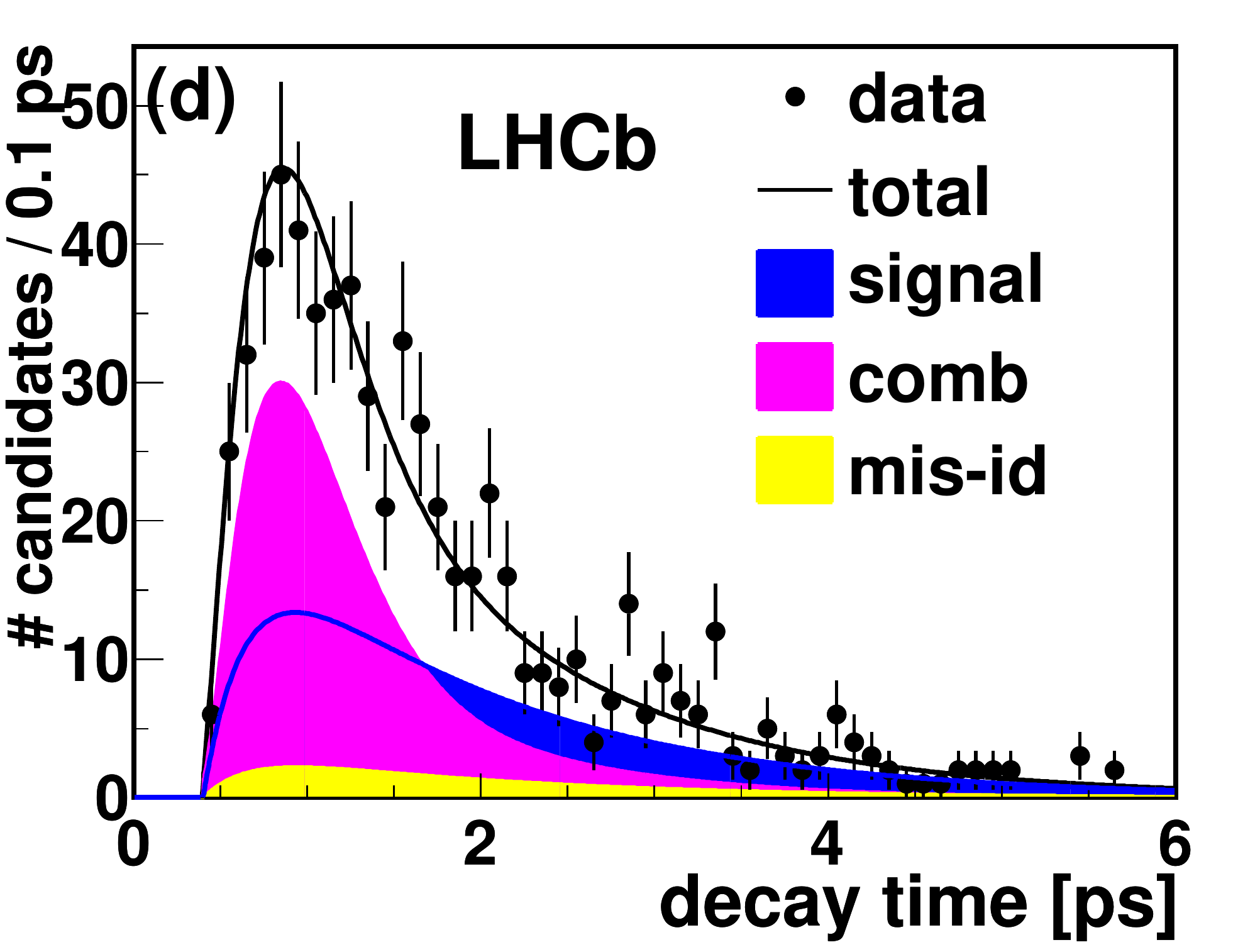}\\
        \caption{Decay time distributions for a) \BsDphipipi, b) \BsDKStKpi, c)
          \BsDKKpipi and d)  \BsDthreepi candidates.
          The data and the fit projection are from a mass range of $\pm$ 3$\sigma_m$ around the reconstructed \Bs
          mass. The abbreviations for the various fit components are introduced in Fig.\ \ref{fig:Bsmass}.}
    \label{fig_ctfit}
  \end{center}
\end{figure}

The shape of the decay time distribution is distorted by trigger and offline
selection criteria which require several particles with large impact parameter
with respect to the primary vertex. This
is accounted  for in the PDF by introducing an acceptance
function $\epsilon(t)$, derived from a full detector simulation. 
Determining $\epsilon(t)$ from simulation is deemed acceptable since it
cancels to first order in the determination of \dms. 
The untagged 
signal decay time PDF becomes

\begin{equation}
\mathcal{P}_t(t | \sigma_t) \propto \left[ \Gs e^{-\Gs\,t} \,
  \cosh \left(\frac{\Delta \Gamma_s}{2}t \right)\,
  \theta(t) \right] \otimes G(t, S_{\sigma_t} \, \sigma_t) \,  \epsilon(t). \label{eq_eff}
\end{equation}
The decay time distributions for the
$b$ decay backgrounds from \Bd and $\Lambda_b$ decays are described
in the same way as that for signal \Bs candidates, using the PDG values for
their lifetimes and \DG=0. 
The shape of the decay time distribution for the combinatorial background is 
described by the sum of two exponential functions multiplied by a second
order polynomial. The parameters of these functions are derived from the high mass sidebands.
Figure \ref{fig_ctfit} illustrates the results of the lifetime
fit.  Within its statistical uncertainty the reconstructed \Bs lifetime
agrees with the PDG value \cite{bib:PDG}.

\section{Flavour tagging}
To determine the flavour of the \Bs~ candidate at production we exploit the
fact that
$b$ quarks are predominantly produced in quark-antiquark pairs.
The quark which is not part of the \Bs~ meson gives rise to an
opposite-side $b$ hadron. 
For opposite-side $b$ hadron decay candidates, the charge of displaced muons, electrons
and kaons and a decay vertex
charge estimate are combined using
a neural network to form a single opposite-side tagging decision.
The tagging decision has a probability to be wrong which is called the mistag probability, $\omega$.
For each event an estimate, $\eta$, of the mistag probability, is determined based upon
topological and kinematic properties of the event, including
the number of primary vertices, the number of tagging particle candidates, the
impact parameter of the tagging particle and of the \Bs candidate with
respect to the primary vertex, and the $p$ and $p_{\mathrm T}$ of
the selected tagging particle  and the \Bs candidate.
The optimization of the tagging algorithms and an initial calibration of
$\eta$ are performed in an independent analysis using
large event samples of $B \rightarrow \mu^+ D^{*-} X$ and \BuJK decays.
More details on the individual tagging algorithms and this calibration procedure
can be found in Ref. \cite{tagging}.

The $B \rightarrow \mu^+ D^{*-} X$ and \BuJK events used in the
optimization and calibration were collected
using different trigger and selection criteria than for the $B^0_s
\rightarrow D_s^-\pi^+$ and $B^0_s \rightarrow D_s^-\pi^+\pi^-\pi^+$ events used in the \dms analysis described here.
As trigger and selection cuts can bias the distributions of the event properties used by the tagging
algorithms, this could result in a biased estimate for the  $B^0_s \rightarrow
D_s^-\pi^+$ and $B^0_s \rightarrow D_s^- \pi^+\pi^-\pi^+$ events.
Therefore, a re-calibration is performed using a sample of 6,000 \BdToDpi events,
which have a similar topology to the $B^0_s \rightarrow D_s^-\pi^+$ and $B^0_s
\rightarrow D_s^-\pi^+\pi^-\pi^+$ events,
and were collected using the same trigger and similar selection cuts.
This event sample is used to perform a measurement of the \Bd-\Bdb flavour
oscillation using a very similar method to that described here.
In that measurement the true event mistag probability, $\omega$, is
parameterized as a linear function of $\eta$ using the relationship
$\omega(\eta) = a + b \times (\eta - \langle \eta \rangle)$,
where $\langle \eta \rangle = 0.3276$ is the mean of the distribution of the $\eta$ values
obtained from the initial tagger optimization.
The parameters $a = 0.311 \pm 0.022$ and $b = 0.61 \pm 0.25$ are determined as
part of the maximum likelihood fit of the
\Bd-\Bdb oscillation signal and found to be consistent with the original
calibration.  
As a by-product of this re-calibration procedure the \Bd-\Bdb oscillation frequency is measured.
The resulting value of
\dmd~=~0.499~$\pm$~0.032~(stat)~$\pm$~0.003~(syst)~ps$^{-1}$, though
statistically less precise, is in
good agreement with the PDG value of  \dmd~= 0.507 $\pm$ 0.004 ps$^{-1}$ \cite{bib:PDG} and provides a valuable cross
check of the procedure.

The statistical power of the tagging is determined by the ``effective''
tagging efficiency for signal events and is defined as

\begin{equation}
\epsilon_{\rm{eff}} = \epsilon_{\rm s} \times \frac{1}{\sum_i{W_i}} \sum_i
(1-2\omega(\eta_i))^2 \times W_i,
\end{equation}
where the signal tagging efficiency $\epsilon_{\rm{s}}$ is a free parameter in
the fit of the oscillation frequency described in the next section. $W_i$ is the
probability for being a signal event as determined by the invariant mass and
decay time PDFs.
The index $i$ runs over all \Bs candidates.

\section{Measurement of the oscillation frequency}
To determine the  oscillation frequency, \dms, the decay time PDF 
for signal candidates 
with tagging information is modified in the following way:
\begin{eqnarray}
 \mathcal{P}_{t}(t, q | \sigma_t, \eta) &\propto& 
\left \{ \Gs e^{-\Gs \, t} \, \frac{1}{2}\left[ \cosh \left( \frac{\Delta \Gamma_s}{2}
t \right) + q \left [1-2\omega(\eta) \right ] \cos(\Delta m_st) \right] \, \theta(t)
  \right \} \nonumber \\
 & & \otimes~ G (t, S_{\sigma_t} \, \sigma_t) \, \epsilon(t) \, \epsilon_{\rm{s}}. 
\end{eqnarray} 
The decay time PDF for
untagged signal events is given by Eq.~(\ref{eq_eff}) multiplied by an additional factor
$(1-\epsilon_{\rm{s}})$.
The calibration parameters $a$ and $b$ of the mistag probability
$\omega(\eta)$ are identical for all signal and $b$
decay background components. Within Gaussian constraints they are set to the values found in the
calibration described in the previous section. The signal tagging efficiency
$\epsilon_{\rm{s}}$ for the \BsDpi and \BsDthreepi
modes are two separate parameters in the fit. The same values of
$\epsilon_{\rm{s}}$ are however used for signal and $b$ decay background components in each of these two categories.
In the description of the combinatorial background a separate parameter
for the tagging efficiency is introduced for each of the four modes. In
addition, tagging asymmetry parameters are introduced in the PDFs for the
combinatorial background, to allow for a different number of
events tagged as \Bs or \Bsb in each mode. As expected
the fit results for these asymmetries are compatible with zero.

\begin{figure}[tb]
  \begin{center}
        \includegraphics[angle=0, width=0.48\textwidth]{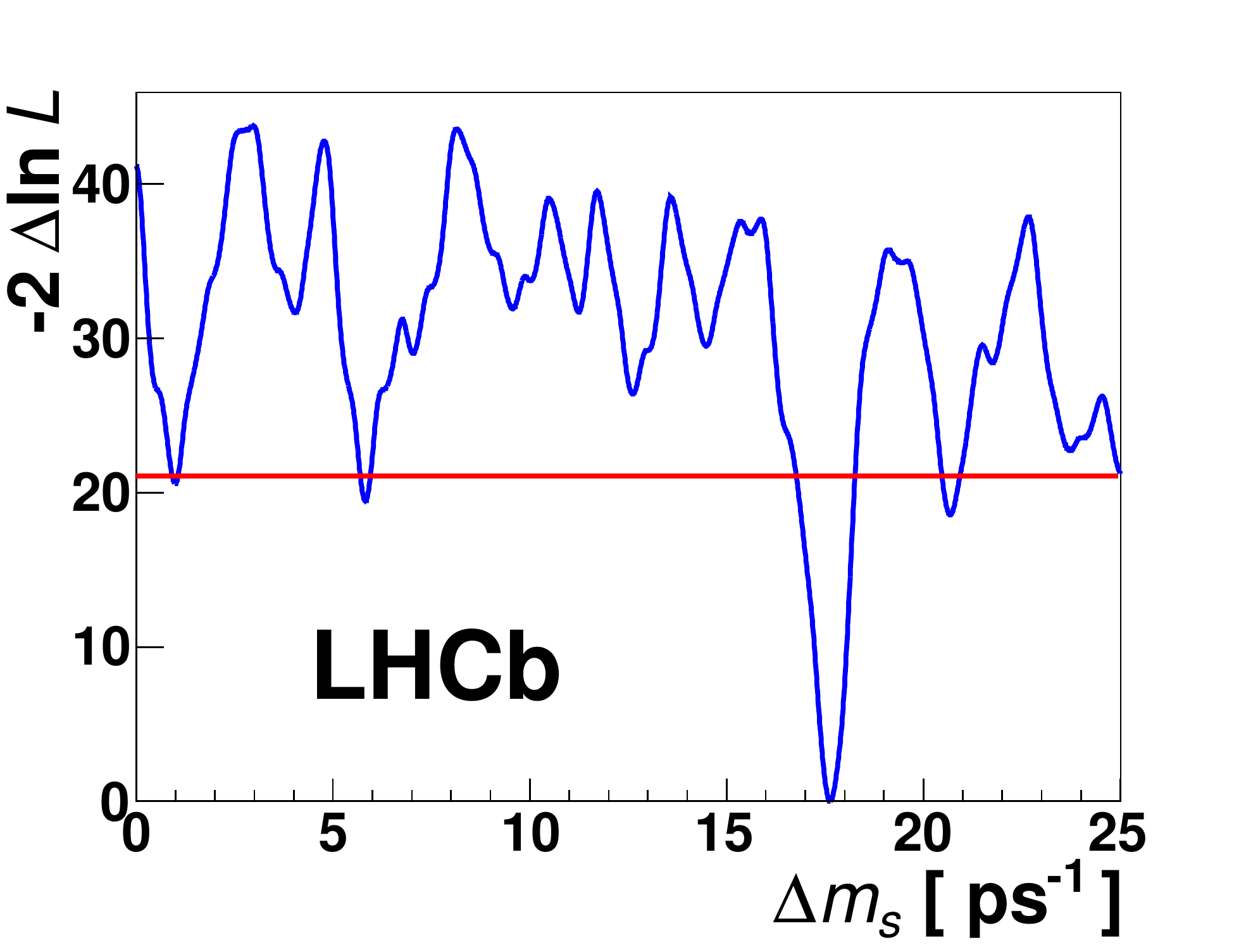}
                \caption{Likelihood scan for \dms in the range
                  [0.0, 25.0] ps$^{-1}$. The line at $-2\Delta \ln L$ = 20.9
          indicates the value in the limit  \dms =
          $\infty$.}
    \label{fig_lik}
  \end{center}
\end{figure}

The fit for the oscillation frequency \dms is performed
simultaneously to all four \Bs decay modes and gives  \dms = 17.63 $\pm$ 0.11
ps$^{-1}$ (statistical uncertainty only).  
Signal tagging efficiencies of $\epsilon_{\rm s}$ = (23.6 $\pm$ 1.3)\,\%
and $\epsilon_{\rm s}$ = (17.6 $\pm$ 3.2)\,\%
are found for the \BsDpi and \BsDthreepi modes, respectively. 
The combined effective tagging efficiency for all four modes 
is $\epsilon_{\rm{eff}}$ = (3.8~$\pm$~2.1)\,\%.
The 
likelihood profile as a function of the assumed oscillation frequency \dms is shown
in Fig.~\ref{fig_lik}.  The statistical significance of the
signal is evaluated to be 4.6\,$\sigma$ by comparing the likelihood value at
the minimum of the fit with that found in the limit \dms = $\infty$.  

\begin{figure}[tb]
  \begin{center}
        \includegraphics[angle=0,          width=0.48\textwidth]{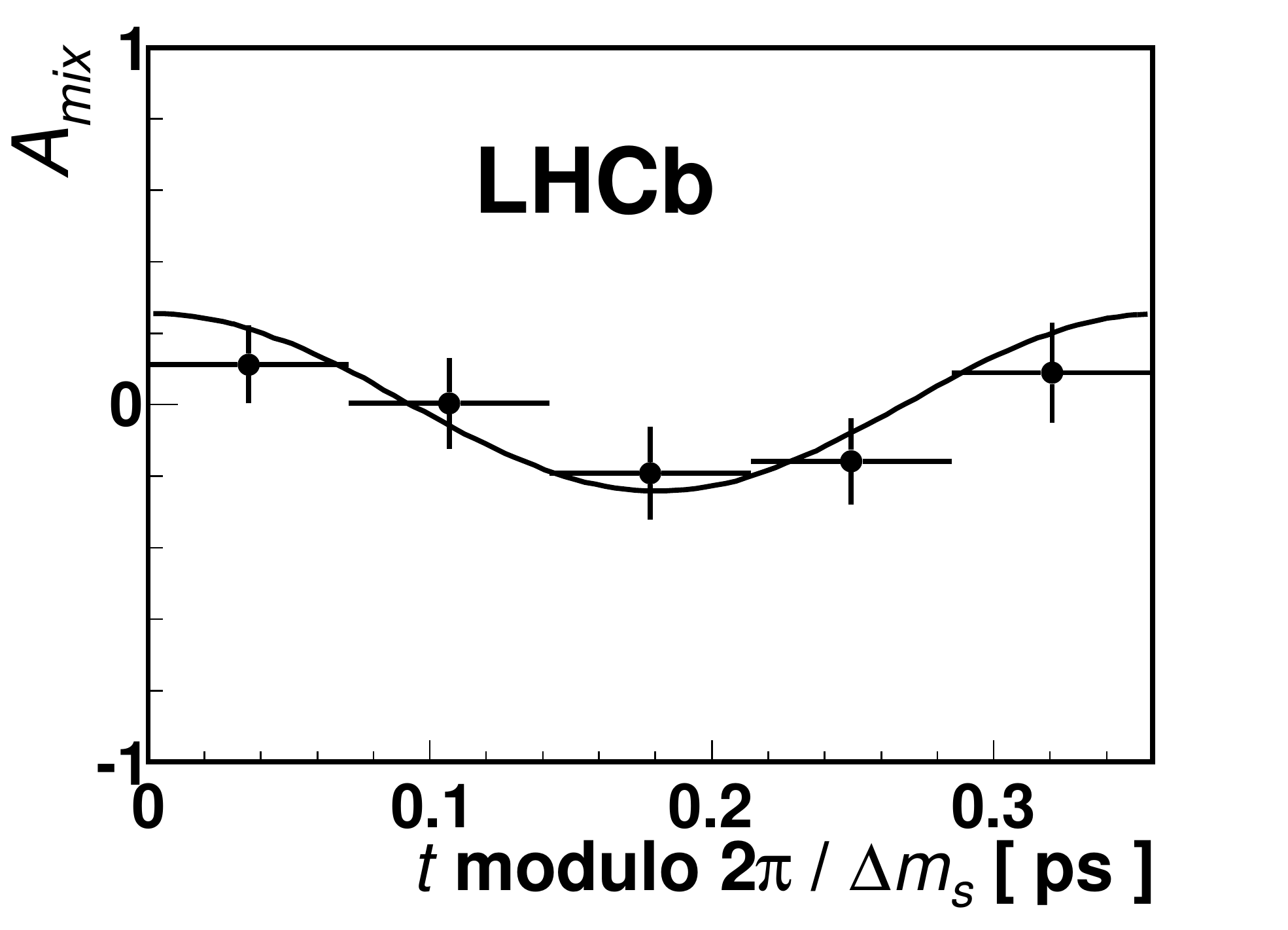}\\
        \caption{Measured asymmetry for \Bs candidates in bins  of the decay
          time $t$ modulo
          $2\pi / \dms$. The projection of  the likelihood fit is superimposed. }
    \label{fig_sens}
  \end{center}
\end{figure}

To illustrate the oscillation pattern, we define the time dependent mixing asymmetry as

\begin{equation}
A_{\rm{mix}}(t) = \frac{N^+(t) - N^-(t)}{N^+(t) + N^-(t)} \\
\label{eq:asym}
\end{equation}
where $N^+(t)$ and $N^-(t)$ are the number of
background subtracted \Bs signal candidates with a given decay time $t$ and 
tagging decision $+1$ and $-1$, respectively.
Note, that this definition of the asymmetry does not include any information on the mistag
probabilities and therefore does not use the full information of the likelihood fit. 
Despite the limited size of the sample, the oscillation pattern is
clearly visible when the asymmetry is plotted in bins of the decay time 
modulo $2\pi/$\dms (Fig.\
\ref{fig_sens}). In an ideal scenario of
perfect tagging and perfect decay time resolution the amplitude of this
oscillation would be 1.0. The observed amplitude is reduced due to the
performance of the tagging algorithm by a
factor 0.41. Another reduction of 0.65 occurs due to the limited decay time resolution.

\section{Systematic uncertainties}
The
dominant source of systematic uncertainty is due to the knowledge of the
absolute decay time scale of the experiment.
This uncertainty is dominated by the knowledge of the $z$ scale. A relative
uncertainty of 0.1\% on the $z$  scale and thus on the decay length is
assigned based on comparisons of detector surveys and a software alignment
using reconstructed tracks. This leads to a systematic uncertainty of
0.018 ps$^{-1}$ on \dms. A second contribution to the decay time scale is due
to the momentum scale of the experiment. From an independent analysis of the mass scale using various
known resonances an uncertainty of the uncalibrated momentum scale of
less than 0.1\% is estimated. This uncertainty partially cancels as it enters 
both the reconstructed \Bs mass and the \Bs momentum. 
The resulting relative uncertainty on the decay time is 0.02\%,
which translates to an absolute systematic uncertainty of 0.004 ps$^{-1}$ on
\dms. 

The next largest systematic uncertainty is related to the description of the
combinatorial background in the fit to the mass spectra. It is evaluated by
replacing the exponential function by a first order polynomial. 
Based on the shift in the value obtained for \dms, a systematic uncertainty of 
0.010 ps$^{-1}$ is assigned. Finally, based on variations of the decay time resolution scale factor
$S_{\sigma_t}$ within its estimated uncertainty from 1.2 to 1.4, a systematic
uncertainty of 0.006
ps$^{-1}$ is assigned on \dms. 
These contributions to the systematic uncertainty on \dms are summarized in
Table\ \ref{tab_sum_syst}.  

Various other possible sources of systematic effects have been studied, such
as the decay time resolution model, the decay time
acceptance, releasing parameters of the invariant mass and decay time PDF in the mixing
fit, different parameterizations of the invariant mass  of the $b$ decay backgrounds and variations of the
value of \DGs.
They are found to be negligible. 

\begin{table}
\begin{center}
\caption{\label{tab_sum_syst} Summary of the systematic uncertainties on
  \dms. The total systematic uncertainty is defined as the quadratic
  sum of the individual components.}
\vspace{0.1cm}

\begin{tabular}{|l|c|}
\hline
Source & Uncertainty [ps$^{-1}$] \\ \hline
Momentum scale                      & 0.004\\
$z$ scale                           & 0.018 \\
Comb. background mass  shape        & 0.010\\
Decay time resolution	            & 0.006\\
\hline
Total systematic uncertainty   & 0.022\\  \hline
\end{tabular}
\end{center}
\end{table}

\section{Conclusion}
A measurement of the \Bs-\Bsb oscillation frequency \dms is performed using
\BsDpi and \BsDthreepi decays collected in 36 pb$^{-1}$ of $pp$ collisions at
$\sqrt{s}$ = 7 TeV in 2010.
The result is found to be
 
\begin{equation}
\dms = 17.63 \pm 0.11 \mathrm{~(stat)} \pm 0.02 \mathrm{~(syst)~ps}^{-1}.
\end{equation}
This is in good agreement with the previous best measurement of \dms = 17.77 $\pm$ 0.10~(stat) $\pm$
0.07 (sys) ps$^{-1}$, reported by the CDF collaboration
\cite{bib:bsmixing}. As a by product of the analysis we also determine a value
for the \Bd-\Bd oscillation frequency \dmd~=~0.499~$\pm$~0.032~(stat)~$\pm$~0.003~(syst)~ps$^{-1}$.
Our results are completely dominated by statistical uncertainties and thus
significant improvements are expected with larger data sets.

\addcontentsline{toc}{section}{References}

\section*{Acknowledgements}

\noindent We express our gratitude to our colleagues in the CERN accelerator
departments for the excellent performance of the LHC. We thank the
technical and administrative staff at CERN and at the LHCb institutes,
and acknowledge support from the National Agencies: CAPES, CNPq,
FAPERJ and FINEP (Brazil); CERN; NSFC (China); CNRS/IN2P3 (France);
BMBF, DFG, HGF and MPG (Germany); SFI (Ireland); INFN (Italy); FOM and
NWO (The Netherlands); SCSR (Poland); ANCS (Romania); MinES of Russia and
Rosatom (Russia); MICINN, XuntaGal and GENCAT (Spain); SNSF and SER
(Switzerland); NAS Ukraine (Ukraine); STFC (United Kingdom); NSF
(USA). We also acknowledge the support received from the ERC under FP7
and the Region Auvergne.


\end{document}